\documentclass[aps,pra,reprint,superscriptaddress,nofootinbib]{revtex4-1}
 
\usepackage{bm}
\usepackage[retainorgcmds]{IEEEtrantools}
\usepackage{graphicx}
\usepackage{mathrsfs}
\usepackage{amsmath}
\usepackage{amsfonts}
\usepackage{amssymb}
\usepackage{color}
\usepackage{times,txfonts}
\usepackage{nicefrac}
\usepackage{ragged2e}
\usepackage{tikz}
\usepackage{enumerate}
\usepackage{mathtools}
\usepackage{cancel}
\usepackage[normalem]{ulem}

\usepackage[colorlinks=true,linkcolor=blue,urlcolor=blue,citecolor=blue,pdfusetitle]{hyperref}

\usepackage{blindtext}

\definecolor{cadmiumgreen}{HTML}{097969}

\newcommand{\victor}[1]{\textcolor{black}{{#1}}}

\newcommand{\victorr}[1]{\textcolor{black}{{#1}}}

\begin{document}

\title{Invisible extended Unruh-DeWitt detector}
\date{\today}
\author{Victor Hugo M. Ramos}
\email{vhmarques@usp.br}
\affiliation{Instituto de F\'isica da Universidade de S\~ao Paulo,  05314-970 S\~ao Paulo, Brazil.}

\author{Jo\~ao Paulo  M.  Pitelli}
\email[]{pitelli@unicamp.br}
\affiliation{Departamento de Matem\'atica Aplicada, Universidade Estadual de Campinas,
13083-859 Campinas, S\~ao Paulo, Brazil}%

\author{João C. A. Barata}
\email{jbarata@if.usp.br}
\affiliation{Instituto de F\'isica da Universidade de S\~ao Paulo,  05314-970 S\~ao Paulo, Brazil.}

\begin{abstract}

We develop a localized particle detector model formulated as a massive quantum field on Minkowski spacetime with the spatial origin excised. To render the problem well-posed at the puncture, we impose boundary conditions at the excised point, which we take to be of Robin type. This setup yields a discrete sector, given by bound-state solutions of the radial equation with real, positive frequencies, which characterizes the detector. We construct the full two-point function and show its decomposition into (i) the discrete radial bound-state sector, (ii) the boundary condition modified continuous sector, and (iii) the unmodified Dirichlet sector. We then compute the detector field’s stress-energy tensor and prove its covariant conservation. For the specific localized modes in this setup, the discrete-sector contribution cancels in the complete stress-energy tensor, leaving only boundary-condition induced terms. Notably, the discrete modes crucial to localized field-based detectors emerge naturally from the boundary conditions, without {\it ad hoc} confining potentials, providing a \victor{covariant} framework that extends the traditional Unruh-DeWitt paradigm.  This mechanism is not restricted to Minkowski spacetime; the same construction can be applied to massive fields on backgrounds with naked singularities, such as conical and global monopole spacetimes, offering a unified route to detector localization in a broad class of geometries.

\end{abstract}

\maketitle{}

\section{Introduction}

Particle detectors play a fundamental role in quantum field theory as operational tools for probing field states and extracting physical information from quantum systems. The theoretical study of detector models has been primarily dominated by the Unruh-DeWitt (UDW) approach \cite{unruh1976notes,dewitt1979general}, where pointlike detectors are modeled as two-level quantum systems coupled locally to quantum fields. This framework has proven invaluable for understanding phenomena such as the Unruh effect~\cite{crispino2008unruh}, Hawking radiation, and entanglement harvesting protocols \cite{liu2022does,APozas-Kerstjens,Foo}. However, despite its widespread success, the UDW model suffers from certain conceptual limitations. The detector itself is described nonrelativistically, and a breakdown of causality and covariance is reveled at length scales smaller than the detector size \cite{de2021relativistic,martin2015causality,martin2021broken,de2023causality}. 

In recent years, there has been growing interest in developing \victor{covariant} detector models based on quantum field theory. One approach involves the Fewster-Verch (FV) measurement framework \cite{fewster2020quantum,fewster2020generally}, which provides a rigorous scheme for measurements in algebraic quantum field theory through dynamical coupling between a system field and a probe field within bounded spacetime regions. While the FV framework offers mathematical rigor and resolves covariance and causality issues in quantum measurements, it is not yet clear how to adapt this framework to concretely model spatially localized detector systems with prescribed internal structure, particularly when one aims at explicit mode decompositions and calculable response functions. A particularly promising alternative approach  involves the use of localized quantum field theories, where the detector degrees of freedom are themselves described by quantum fields confined to specific spatial regions through confining potentials \cite{perche2024particle, bruno1, bruno2}. This approach addresses several shortcomings of traditional UDW detector models by providing a manifestly covariant description that naturally incorporates the detector's internal structure and backreaction effects, while remaining computationally accessible for explicit calculations of detector responses and correlations.

In this work, we present a \victor{covariant} detector model formulated as a quantum field in Minkowski spacetime with the spatial origin excised, in which Robin-type boundary conditions at the removed point naturally generate a discrete sector associated with radial bound modes of real, positive frequency. This spectral structure provides localized detector degrees of freedom without introducing {\it ad hoc} confining potentials, reconciling locality and covariance and offering an operationally transparent avenue for comparison with Unruh-DeWitt-type detectors. We construct the full two-point function and show its decomposition into three physically distinct contributions: (i) the discrete bound-state sector, (ii) the spherically symmetric continuous sector modified by the Robin boundary condition, and (iii) the unmodified Dirichlet sector. From this decomposition, we compute the renormalized stress-energy tensor of the detector field by Hadamard subtraction and demonstrate its covariant conservation. A notable result is that the discrete-sector contribution cancels exactly, in the complete observable, against the corresponding pole in the modified continuum, so that the remaining effects are purely induced by the boundary condition. Thus, the discrete modes that characterize localized detectors emerge intrinsically from the self-adjoint extensions of the radial operator, while the semiclassical gravitational response is entirely encoded in boundary terms.

 We remark that punctured Minkowski spacetime serves as a convenient toy model for more physically relevant scenarios. In particular, the same construction extends naturally to spacetimes with naked singularities, provided that the boundary condition induces a localized sector. This occurs, for instance, in the 
(1+2)-dimensional spacetime of a point source~\cite{kay} (the conical spacetime) and in the spacetime surrounding a global monopole~\cite{pitelliglobalmonpole}.

The paper is organized as follows. In Sec. \ref{sec:II} we establish the dynamics of the scalar field in punctured Minkowski spacetime, classify the relevant self-adjoint extensions, and identify the Robin conditions that generate the spherically symmetric bound mode. In Sec. \ref{sec:III} we construct the mode expansion, derive the two-point function, and make explicit its decomposition into discrete, modified continuous, and Dirichlet sectors. We then formulate the coupling of the localized detector field to an external scalar field and discuss the leading-order correspondence with the Unruh-DeWitt model. In Sec. \ref{sec:IIIB} we compute the renormalized $\langle \Psi^2 \rangle$ and the stress–energy tensor via Hadamard subtraction, demonstrating covariant conservation and the absence of a net contribution from the discrete mode to the full tensor. We conclude in Sec. \ref{concluding remarks} with remarks on natural extensions, including couplings with compact support, nonvacuum states, and generalizations to curved spacetimes.

\section{Field Theory in Punctured Minkowski}\label{sec:II}

We consider a real massive scalar field $\Psi$ propagating in four-dimensional Minkowski spacetime in spherical coordinates with the spatial point $r=0$ removed. This “punctured” Minkowski spacetime allows a broader class of admissible boundary conditions at the excised point. Our objective is to analyze how these boundary conditions affect the spectrum of the field and, in particular, to show that positive-frequency discrete modes, corresponding to bound states in the radial equation, can emerge.

\subsection{Scalar field dynamics and boundary conditions}

To analyze the dynamics of a scalar field, we now study the solutions to the Klein-Gordon equation. We employ standard spherical coordinates \((t, r, \theta, \phi)\), with \(r > 0\) and the origin \(r = 0\) removed. The spacetime line element takes the form
\begin{align}
ds^{2} = -dt^{2} + dr^{2} + r^{2} \left( d\theta^{2} + \sin^{2}\theta\, d\phi^{2} \right),
\end{align}

The field satisfies the Klein-Gordon equation,
\begin{align}
\left( -\frac{\partial^{2}}{\partial t^{2}} + \frac{\partial^{2}}{\partial r^{2}} + \frac{2}{r} \frac{\partial}{\partial r} + \frac{1}{r^{2}} \Delta_{\mathbb{S}^2} - m_0^2 \right)\Psi(t, r, \theta, \phi) = 0,
\end{align}
where \(\Delta_{\mathbb{S}^2}\) denotes the Laplacian on the unit 2-sphere,
\begin{align}
\Delta_{\mathbb{S}^2} = \frac{1}{\sin\theta} \frac{\partial}{\partial \theta} \left( \sin\theta \frac{\partial}{\partial \theta} \right) + \frac{1}{\sin^2\theta} \frac{\partial^2}{\partial \phi^2}.
\end{align} We seek separable solutions of the form
\begin{align}\label{ansatz}
\Psi_{\omega \ell m}(t, r, \theta, \phi) = e^{-i\omega t} R_{\omega\ell}(r) Y_{\ell m}(\theta, \phi),
\end{align}
where \(Y_{\ell m}(\theta, \phi)\) are the standard spherical harmonics,
\begin{align}
\Delta_{\mathbb{S}^2} Y_{\ell m} = -\ell(\ell+1) Y_{\ell m}, \quad \ell \in \mathbb{N}_0, \quad m \in \{-\ell, \dots, \ell\}.
\end{align} Inserting this ansatz into the field equation yields the following radial equation for \(u_{\omega  \ell } (r) = r R_{\omega \ell}(r)\):
\begin{align}\label{calogero}
Au_{\omega \ell} \coloneqq \left(-\frac{d^{2}}{dr^{2}}+\frac{\ell\left(\ell+1\right)}{r^{2}}\right)u_{\omega\ell}=p^{2}u_{\omega\ell},
\end{align} where $p^2 = \omega^2 - m_0^2$ is the spectral parameter. Therefore, the analysis of the scalar field reduces to a Calogero-type quantum mechanical problem on the positive real axis. Since the associated differential operator is singular at the origin, a well-defined and unitary time evolution requires that the operator be self-adjoint. This, in turn, demands the specification of suitable boundary conditions at $r=0$, which correspond to choosing a particular self-adjoint extension of the operator $A$. A thorough classification of the self-adjoint extensions of the Calogero operator \( A_C = -\frac{d^2}{dr^2} + \frac{a}{r^2} \) is presented in \cite{gitman2010self}, where the domains associated with each extension characterize the most general square-integrable solutions consistent with the boundary behavior at \( r = 0 \). The classification depends on whether \( -\frac{1}{4} < a < \frac{3}{4} \) or \( a \geq \frac{3}{4} \). In our case, identifying \( a = \ell(\ell + 1) \), this corresponds to analyzing separately the cases \( \ell = 0 \) and \( \ell > 0 \). We shall elaborate on each of these cases.

\subsubsection{$\ell >0$}

In this case, which corresponds to the nonspherically symmetric modes, the operator $A$ defined on the domain of test functions $C^\infty_0 (\mathbb{R}_+)$ is essentially self-adjoint. Therefore, only the Dirichlet boundary condition is admissible at $r=0$ and the eigenfunctions of $A$ with eigenvalue $p^2>0$ are \begin{align}\label{eq:RadialSolutionDirich}
    u_{\omega \ell }\left(r\right) = \sqrt{\dfrac{pr}{2}}J_{\ell + 1/2}\left( pr \right),
\end{align} where $J_{\ell + 1/2}$ are the Bessel functions of order $\ell +1/2$. Furthermore, these radial solutions are normalized in $L^2(\mathbb{R}_+,r^2 dr)$, implying the normalization of (\ref{ansatz}) with respect to the Klein-Gordon product, i.e. \begin{align}\label{kg-normalization}
    \left( \Psi_{\omega \ell m}, \Psi_{\omega' \ell' m'} \right)_{K.G.}=\delta(\omega-\omega')\delta_{\ell \ell'} \delta_{mm'}.
\end{align}

\subsubsection{$\ell = 0$}

In this sector, the operator $A$ is not self-adjoint in the domain $C^\infty_0(\mathbb{R}_+)$, but there is a one-parameter $U(1)$-family of self-adjoint extensions. In fact, for each parameter $\beta \in \mathbb{R}$, there is a self-adjoint extension $A_\beta$, whose domain includes solutions of the eigenvalue problem (\ref{calogero}) satisfying~\cite{pitelliglobalmonpole} \begin{align}\label{boundarycondition}
    \lim_{r \to 0^+}\left[ r R_{\omega}^{\left(\beta\right)}\left(r\right)+\beta\left( r R_{\omega}^{\left(\beta\right)}\left(r\right)\right)'  \right]=0,
\end{align} where the prime denotes the derivative with respect to $r$. Hence, the self-adjoint extensions are characterized by Robin boundary conditions at the excised point. Assuming $p^2>0$, the general solution takes the form \begin{align}\label{eq:RadialSolutionRobin}
    R_{\omega}^{\left(\beta\right)}\left(r\right) =\frac{p\beta\cos\left(p r\right)-\sin\left(p r\right)}{2\pi r\sqrt{p\left(1+p^{2}\beta^{2}\right)}}.
\end{align} This solution is normalized with respect to (\ref{kg-normalization}), and it encodes the choice of boundary condition through the parameter $\beta$. 

While the solution discussed above correspond to the continuous part of the spectrum, the possibility of discrete modes also arises due to boundary conditions. These are associated with negative values of $p^2=-\mu^2$, \victorr{where $\mu^2 = 1/\beta^2$ is fixed by the boundary condition~(\ref{boundarycondition}) and it is related to the bound mode positive frequency $\omega_b \equiv \sqrt{m_0^2-1/\beta^2}$ given that $\beta m_0 > 1$ (in the following analysis, the latter will always be assumed)}. \victorr{Hence,} the general solutions satisfying Robin boundary conditions at $r=0$ are of the form \begin{align}
    R_{\text{bound}}^{(\beta)}\left(r\right)=\sqrt{\frac{1}{4\pi\beta\omega_{\text{b}}}}\frac{e^{-\frac{r}{\beta}}}{r}.
    \label{squareint}
\end{align} \victorr{Given the singular character of the punctured Minkowski spacetime, the boundary condition (\ref{boundarycondition}) models the interaction between the field and the excised point. This interaction introduces a new physical scale $\beta$ to the system which is responsible for the existence of a localized mode. The choice for this particular interaction (characterized by Robin boundary conditions) is due to the fact that the dynamics of the resulting field is physically sensible \cite{WaldII}.}

\subsection{The quantum field}\label{sec:detectorquantumtheory}
The existence of bound states in the radial equation directly translates into the presence of discrete modes in the quantum field expansion, which takes the form \begin{multline}
    \hat{\Psi}_\beta(x)=\Psi_{\textrm{bound}}(x) \hat{a}_\textrm{bound}+\Psi_\textrm{bound}^*(x)\hat{a}^\dagger_\textrm{bound} \\
    +\sum_{\ell=1}^\infty \sum_{m=-\ell}^\ell \int_{m_0}^\infty d\omega \left( \Psi_{\omega \ell m}(x)\hat{b}_{\omega \ell m}+ \Psi_{\omega \ell m}^*(x)\hat{b}^\dagger_{\omega \ell m}\right), 
\end{multline} where $\Psi_{\omega \ell m}(x)$ is given by Eq.(\ref{ansatz}) with the respective radial solutions (\ref{eq:RadialSolutionDirich}) and (\ref{eq:RadialSolutionRobin}). Additionally, the discrete mode solution takes the form \begin{align}\label{eq:DiscreteMode}
    \Psi_{\textrm{bound}}(x) = \dfrac{e^{-i\omega_b t}}{\sqrt{4\pi}}R_{\textrm{bound}}(r),
\end{align} and the creation and annihilation operators $\hat{a}_{\textrm{bound}}$, $\hat{a}^\dagger_{\textrm{bound}}$, $\hat{b}_{\omega \ell m}$, $\hat{b}^\dagger_{\omega \ell m}$ satisfy the canonical commutation relations \begin{align}
    \left[ \hat{a}_{\textrm{bound}},\hat{a}_{\textrm{bound}}^\dagger \right] & = 1, \\
    \left[ \hat{b}_{\omega \ell m},\hat{b}_{\omega' \ell' m'}^\dagger \right] & = \delta(\omega-\omega') \delta_{\ell\ell'}\delta_{mm'}.
\end{align} The annihilation operators define the vacuum state $|0\rangle$ via $\hat{a}_{\textrm{bound}} | 0 \rangle = 0 $ and $\hat{b}_{\omega \ell m}| 0\rangle = 0 $ for all $\omega$, $\ell$, $m$. They additionally specify no excitation states of each sector, i.e. the state $| 0_{\textrm{bound}} \rangle$ such that $\hat{a}_{\textrm{bound}}| 0_{\textrm{bound}} \rangle =0 $ defines the vacuum of the discrete mode, and $| 0_{\textrm{cont}} \rangle$ such that $\hat{b}_{\omega \ell m}| 0_\textrm{cont}\rangle = 0 $ for all $\omega$, $\ell$, $m$ defines the unexcited state associated to the continuum modes. Hence, the vacuum decomposes to $| 0 \rangle = | 0_{\textrm{bound}} \rangle \otimes | 0_{\textrm{cont}} \rangle $ and the Fock space reads \begin{align}
    \mathcal{F} = H_{\textrm{bound}}\otimes \mathcal{F}_{\textrm{cont}},
\end{align} where each space is expressed as \begin{align}
    H_{\textrm{bound}}  = \textrm{span} \left( \left\{ \left( \hat{a}_{\textrm{bound}}^\dagger \right)^n | 0_{\textrm{bound}} \rangle:\ n\in \mathbb{N} \right\} \right), \\
    \mathcal{F}_{\textrm{cont}}  = \textrm{span} \left( \left\{ \left( \hat{b}_{\omega \ell m}^\dagger \right)^n | 0_{\textrm{cont}} \rangle:\ n\in \mathbb{N},\ (\omega,\ell,m)\in \Omega \right\} \right),
\end{align} with the space of parameters given by $\Omega = \{ (\omega, \ell, m): \omega\in (m_0,\infty),\ell\in\mathbb{N}_0,m\in\{-\ell,...,\ell\}\}$.

\section{The Detector}\label{sec:III}

In this section we review the modeling of a particle detector by a localized quantum field $\hat{\phi}_D(x)$ presenting a discrete set of modes, and acting as a probe for a free massless Klein-Gordon field $\hat{\phi}(x)$. By identifying the detector’s field degrees of freedom with the field $\hat{\Psi}_\beta$ prescribed in Sec. \ref{sec:II}, which is naturally equipped with a discrete mode sector, we obtain a detector model that is localized. This setting allows us to investigate the detector’s vacuum polarization $\langle \hat{\Psi}^2_\beta \rangle$ and stress-energy content from first principles, taking into account the full quantum field-theoretic structure of $\hat{\Psi}_\beta$.

\subsection{Localized detector}

The development of particle detector models in quantum field theory has evolved significantly since the introduction of the paradigmatic Unruh-DeWitt detector \cite{unruh1976notes,dewitt1979general}. Traditionally, it describes a detector as an harmonic oscillator of frequency $\Omega$ that couples locally to a quantum field $\hat{\phi}$ through an interaction prescribed by the Hamiltonian \begin{align}\label{eq:Unruh-DeWittHamiltonian}
    \hat{h}_I = \lambda \Lambda(x)\left( e^{-i \Omega t} \hat{a} + e^{i \Omega t} \hat{a}^\dagger \right)\hat{\phi}(x).
\end{align} Within this framework, the oscillator system is prescribed by the creation and annihilation operators $\hat{a}$, $\hat{a}^\dagger$ which couples linearly with the local quantum field by the interaction parameter $\lambda$ in a spacetime region specified by the smearing function $\Lambda(x)$. The profile of the later guarantees the localization of the model interaction. 

While the Unruh-DeWitt model provides a conceptually simple prescription for particle detection, it relies on introducing a nonrelativistic degree of freedom. An alternative and \victor{covariant} approach consists in modeling the detector itself as a localized quantum field coupled to the field of interest. In this framework, the detector’s degrees of freedom are those of a quantum field with discrete spectrum that interacts locally with a free field. This formulation inherently ensures covariance and allows for the construction of observables such as the detector’s stress-energy tensor \cite{perche2025stress}. Interestingly, it has been shown that localized quantum field detectors reproduce the same response as Unruh-DeWitt detectors at leading order in perturbation theory. This equivalence, discussed in detail in \cite{perche2024particle}, provides a deeper interpretation of the Unruh-DeWitt prescription as an effective model arising from a more fundamental field-theoretic description of detectors.

In what follows, we reconsider the analogy between Unruh-DeWitt detectors and localized quantum field detectors, now focusing on the specific field-theoretic degrees of freedom introduced in Sec. \ref{sec:II}. In particular, we interpret the field $\hat{\Psi}_\beta$ as providing a concrete realization of a localized detector. This perspective allows us to establish a direct comparison between the behavior of the field’s bound mode and that of a conventional Unruh-DeWitt detector, with both acting as quantized subsystems locally coupled to an external quantum field. The interaction between the localized detector and an external massless scalar field $\phi$ \victorr{on the punctured Minkowski background} is described by the Lagrangian density \begin{align}
    \mathcal{L} = -\dfrac{1}{2} \partial_\mu \phi_D \partial^\mu \phi_D - \dfrac{1}{2}\partial_\mu \phi \partial^\mu \phi-\dfrac{m_D^2}{2}\phi_D^2-\lambda \zeta(x)\phi_D \phi,
\end{align} where $\lambda$ is the coupling real constant and $\zeta(x)$ is the smearing function such that its profile define the spacetime region where the interaction takes place~\footnote{\victor{The field $\phi$ should, in principle, satisfy arbitrary Robin boundary condition at the excised point as well. If we consider Dirichlet boundary condition, we will be effectively modeling a target field $\phi$ in full Minkowski spacetime interacting with a detector ``centered'' at $r=0$.} }. Hence, in the quantum setup of Sec. \ref{sec:detectorquantumtheory}, we can determine the time evolution of a state $\hat{\rho}_0=\hat{\rho}_D \otimes \hat{\rho}_\phi$, where $\hat{\rho}_\phi$ represents the state of the field $\hat{\phi}(x)$ and $\hat{\rho}_D=\hat{\rho}_{\textrm{cont}}\otimes\hat{\rho}_{\textrm{bound}}$ with $\hat{\rho}_{\textrm{cont}}$ a state in $\mathcal{F}_{\textrm{cont}}$ and $\hat{\rho}_{\textrm{bound}}$ a state in $H_{\textrm{bound}}$. The time evolution follows from the Hamiltonian density, \begin{align}\label{eq:HamiltonianInteraction}
    \mathcal{H}_I = \lambda \zeta(x)\hat{\phi}_D(x)\hat{\phi}(x),
\end{align} which generates a unitary time evolution operator $\hat{U}_I$. In this scenario, we obtain to leading order in $\lambda$ that the final state has the form \cite{perche2024particle}\begin{align} \label{eq:EvolvedState}
    \hat{\rho}= \textrm{tr}_{\phi,\mathcal{F}_{\textrm{cont}}} \left( \hat{U}_I \hat{\rho}_0 \hat{U}_I^\dagger \right).
\end{align} Moreover, by identifying the replacements $\hat{a}\to \hat{a}_{\textrm{bound}}$, $\Lambda(x)e^{- i\Omega t}\to \zeta(x)\Psi_{\textrm{bound}}(x)$, $\Lambda(x)e^{i\Omega t}\to \zeta(x)\Psi_{\textrm{bound}}^* (x)$ in Eq. (\ref{eq:Unruh-DeWittHamiltonian}) and evolving $\hat{\rho}_0$ accordingly, the results of \cite{perche2024particle} confirm that the evolved state matches exactly (\ref{eq:EvolvedState}).

\victor{In fact, having defined the detection quantum field $\hat{\phi}_{D}(x)$, the process of detection is operationally defined as follows. First, the detector field interacts with a separate "target" free field $\hat{\phi}(x)$, which is the system being probed, through the Hamiltonian in Eq.~(\ref{eq:HamiltonianInteraction}). The detector field $\hat{\phi}_{D}(x)$, due to the excised point $r=0$ in its definition, possesses a discrete energy mode (the bound mode). The crucial step, which constitutes the ``detection'', occurs when an observer is assumed to have access to only this single bound mode. To find the state of this accessible mode, the system's evolution is calculated, and then all inaccessible degrees of freedom---namely, the entire target field $\hat{\phi}(x)$ and all other modes of the detector field---are traced out. The resulting final state of this mode, given by Eq.~(\ref{eq:EvolvedState}), contains information about the target field and, to leading order, behaves as an Unruh-DeWitt detector \cite{perche2024particle}.}

These results establish a direct correspondence between the bound-mode sector of a quantum field with Robin boundary conditions and the Unruh-DeWitt detector model. The approach we presented describes a relativistic detector, and provides a clear mechanism through which localized, discrete detector modes arise naturally from the structure of the field and its boundary conditions.

\subsection{Quantum expectation values}\label{sec:IIIB}

In this subsection, we focus on the quantum field-theoretic aspects of the localized detector, specifically analyzing its two-point correlation function and stress-energy tensor \victor{only for the detector model, i.e., without the target field $\hat{\phi}$}. These observables\victor{, while not directly related to the detection procedure,} provide a \victor{straightforward} characterization of the detector's microscopic degrees of freedom and offer insight into both its quantum fluctuations and its backreaction on the surrounding spacetime.

The analysis of the detector’s quantum fluctuations and mean energy-momentum content naturally begins with the explicit form of its two-point correlation function. Accordingly, we now proceed to derive and discuss the Green’s function associated with the detector field, focusing on how it reflects the discrete spectral features introduced previously. In fact, the solutions in Eq. (\ref{ansatz}), with the associated radial solutions (\ref{eq:RadialSolutionDirich}) and (\ref{eq:RadialSolutionRobin}), along with the discrete mode solutions in Eq. (\ref{eq:DiscreteMode}) form a complete set of normalized mode functions, which allows to evaluate the Green's function as \begin{multline}
    G(x,x')= \Psi_{\textrm{bound}}(x) \Psi_{\textrm{bound}}^*(x')\\+\sum_{\ell=0}^\infty \sum_{m=-\ell }^\ell \int_{m_0}^\infty d\omega \ \Psi_{\omega \ell m}(x)\Psi_{\omega \ell m}^*(x').
\end{multline} 

Since the boundary conditions affect exclusively the spherically symmetric sector ($\ell=0$), the Green's function can be naturally decomposed into three distinct contributions: (i) the discrete bound state modes arising from the self-adjoint extension of the radial operator [$G_{\textrm{bound}}(x,x')$], (ii) the continuous  mode in the $\ell=0$ sector modified by the Robin boundary conditions [$G_{\beta}(x,x')$], and (iii) the modes corresponding to Dirichlet boundary condition which remain unaffected by the chosen extension [$G_{\textrm{Dirichlet}}(x,x')$]. This separation aligns with the spectral structure of the problem and facilitates the detailed analysis of each component’s role. Therefore, \begin{align}\label{eq:GreensFunctionDecomposition}
    G(x,x')=G_\textrm{bound}(x,x')+G_{\beta}(x,x')+G_{\textrm{Dirichlet}}(x,x'),
\end{align} where each contribution takes the form \begin{align}
    G_{\text{bound}}\left(x,x'\right)=\frac{1}{4\pi\beta\omega_{\text{b}}}\frac{e^{-\frac{r+r'}{\beta}}e^{-i\omega_{\text{b}}\left(t-t'\right)}}{rr'},
\end{align} \begin{widetext}\begin{align}\label{eq:TPF}
   G_{\beta}\left(x,x'\right)=\frac{1}{4\pi^{2}rr'}\int_{0}^{\infty}dp\left[\frac{e^{-i\sqrt{p^{2}+m_{0}^{2}}\left(t-t'\right)}}{\sqrt{p^{2}+m_{0}^{2}}}\left(\frac{p\beta\left(p\beta\cos\left(p\left(r+r'\right)\right)-\sin\left(p\left(r+r'\right)\right)\right)}{1+p^{2}\beta^{2}}\right)\right],
\end{align} \begin{align}
    G_{\text{Dirichlet}}\left(x,x'\right)=\frac{1}{4\pi^{2}}\sum_{\ell=0}^{\infty}\sum_{m=-\ell}^{\ell}\int_{0}^{\infty}dp\left[\frac{p e^{-i\sqrt{p^{2}+m_{0}^{2}}\left(t-t'\right)}}{\sqrt{p^{2}+m_{0}^{2}}}j_{\ell}\left(p r\right)j_{\ell}\left(p r'\right)Y_{\ell m}\left(\theta,\phi\right)Y_{\ell m}\left(\theta',\phi'\right)\right].
\end{align}\end{widetext} The contribution to the Green’s function arising from the Robin boundary condition smoothly vanishes as $\beta\to 0$, thereby recovering the standard Dirichlet case. In order to make the analytic structure of $G_\beta(x,x')$ more transparent and to facilitate its analysis, it is useful to rewrite it in a form that isolates the pole structure and separates it from the contributions associated with the branch cut of the square root. To this end, we employ an integral contour into the complex $p$-plane. The contour is chosen so as to avoid both the branch cut of the square root in the integrand and the simple poles arising from the rational part of the integrand. This procedure allows us to express the original integral as the sum of the contributions from the enclosed poles and an integral over the branch cut discontinuity. In the latter, additional poles emerge, which are handled using partial fraction decomposition and rewritten via the identity
\begin{equation}
\frac{1}{p \pm \frac{i}{\beta}} = \int_0^\infty ds\, e^{-s\left(p \pm \frac{i}{\beta}\right)}.
\end{equation} The resulting integral leads to the expression
\begin{widetext}\begin{align}\label{eq:GreenBoundaryCondition}
    G_{\beta}\left(x,x'\right)=-\frac{1}{4\pi\beta\omega_{\text{b}}}\frac{e^{-\frac{r+r'}{\beta}}e^{-i\omega_{\text{b}}\Delta t}}{rr'} + \frac{1}{4\beta\pi^{2}rr'}\int_{0}^{\infty}ds\cosh\left(\frac{s}{\beta}\right)\int_{m_{0}}^{\infty}dp\frac{e^{-p\left(s+r+r'\right)}\cosh\left( (t-t') \sqrt{p^{2}-m_{0}^{2}}\right)}{\sqrt{p^{2}-m_{0}^{2}}}.
\end{align}\end{widetext} The first contribution arises from the poles of the integrand and coincides, up to a sign difference, with $G_{\mathrm{bound}}(x,x')$. Consequently, in the full Green's function $G(x,x')$, the discrete mode contribution corresponding to the detector's sector, exactly cancels out. On the other hand, the second contribution is regular and have a well-defined coincidence limit $(t',r')\to(t,r)$ given by \begin{align}
    \left\langle \Psi^{2}\right\rangle _{\beta} = \frac{1}{4\beta\pi^{2}r^{2}}\int_{0}^{\infty}ds\,\cosh\left(\frac{s}{\beta}\right)K_{0}\left(m_{0}\left(s+2r\right)\right),
\end{align} where the integral over $p$ in Eq. (\ref{eq:GreenBoundaryCondition}) is rewritten by a standard integral representation of the Bessel $K_0$.

\begin{figure}[htbp]
    \centering
    \includegraphics[width=0.5\textwidth]{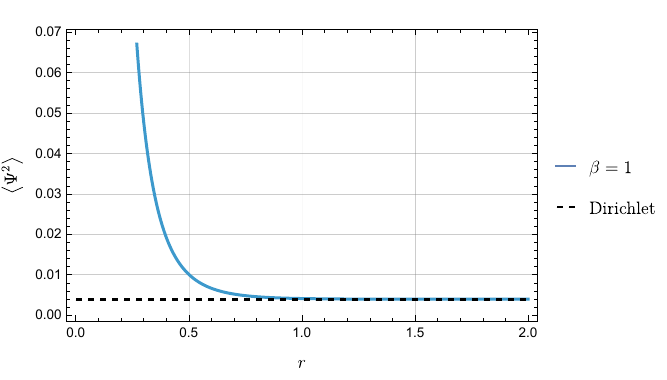}
    \caption{Renormalized vacuum expectation value $\langle \Psi^{2}\rangle$ for $m_{0}=2$ and $\beta=1$. 
    The Dirichlet term acts as a constant contribution, while the $r$-dependence is entirely due to the boundary conditions.}
    \label{fig:Psi2}
\end{figure}

We now address the Dirichlet sector, $G_{\textrm{Dirichlet}}(x,x')$. Physically, it describes the behavior of the quantum field in the region unaffected by the modified boundary, and it coincides with the canonical expression for the Green's function for a massive scalar field propagating in Minkowski spacetime. Indeed, upon application of the addition theorems for both spherical harmonics and Bessel functions, the resulting integral matches the standard representation of the modified Bessel function $K_1$, thus yielding \begin{align}
   G_{\text{Dirichlet}}\left(x,x'\right)= \frac{m_{0}}{4\pi^{2}\sigma}K_{1}\left(m_{0}\sigma\right),
\end{align} where $\sigma = \sqrt{- (t-t')^{2}+r^{2}+r'^{2}-2rr'\cos\gamma}$ is the square root of the geodesic distance, and $\cos \gamma =\cos\theta\cos\theta'+\sin\theta\sin\theta'\cos\left(\phi-\phi'\right)$. Its singular behavior around the null singularity becomes explicit by the following representation \begin{multline}
    G_{\text{Dirichlet}}\left(x,x'\right)=\frac{1}{4\pi^{2}\sigma^{2}}+\ln\left(\sigma\right)\frac{I_{1}\left(m_{0}\sigma\right)}{\sigma}+\\+\frac{m_{0}}{8\pi^{2}}\ln\left(\frac{m_{0}^{2}}{2}\right)\frac{I_{1}\left(m_{0}\sigma\right)}{\sigma}+\\-\frac{m_{0}^2}{16\pi^{2}}\sum_{k=0}^{\infty}\left(\psi\left(k+1\right)+\psi\left(k+2\right)\right)\frac{\left(\frac{1}{2}m_{0}\sigma\right)^{2k}}{k!\left(k+1\right)!},
\end{multline} where $\psi(x)$ is the polygamma function, and the series representation of $K_1 (x)$ was employed \cite{Gradshteyn}. After performing the Hadamard subtraction, the coincidence limit gives the vacuum expectation value of the field square, \begin{align}
    \left\langle \Psi^{2}\right\rangle = \left\langle \Psi^{2}\right\rangle_\beta +\left\langle \Psi^{2}\right\rangle_{\textrm{Dirichlet}},
\end{align} where the Dirichlet contribution takes expected the form \begin{align}
    \left\langle \Psi^{2}\right\rangle_{\textrm{Dirichlet}} = \frac{m_{0}^{2}}{16\pi^{2}}\ln m_{0}^2,
\end{align} vanishing in the $m_0 \to 0$ limit.  To illustrate this result, in Fig. \ref{fig:Psi2} we display the behavior of the renormalized vacuum expectation value $\langle \Psi^{2}\rangle$ for the illustrative choice $m_{0}=2$ and $\beta=1$. In this case the Dirichlet term contributes solely as a constant shift, while all the $r$-dependent behavior originates from the boundary contribution. This provides a clear picture of how the boundary conditions modify the field fluctuations: the boundary-induced term is dominant in the vicinity of the excised point $r=0$, whereas it rapidly decays with increasing $r$, leaving the constant Dirichlet part as the leading contribution at large distances.

In summary, the proposed decomposition clarifies that the Dirichlet contribution alone reproduces the expected local Hadamard behavior prescribed by quantum field theory, while any modifications arising from nontrivial boundary conditions are fully captured by the additional terms. Remarkably, these terms exhibit a cancellation of the radial bound-state contribution by the poles associated with the boundary condition sector, a phenomenon analogous to that observed in the one-dimensional $\delta$-function potential \cite{blinder1988green} and in the Coulomb potential \cite{blinder1981nonrelativistic,hostler1964coulomb}.

\subsubsection{Stress-energy tensor}

A fundamental observable in quantum field theory on curved spacetimes is the expectation value of the stress-energy tensor operator, which characterizes the energy-momentum distribution of quantum fields and serves as the source term in the semiclassical Einstein equations. In the present context, although the nontrivial boundary conditions induce measurable modifications to this tensor, remarkably, the discrete bound state mode provides no contribution to its expectation value. This can be verified directly from the Green's function in Eq.~(\ref{eq:GreensFunctionDecomposition}), where the discrete mode contribution is exactly canceled by the corresponding pole arising from the continuous sector with Robin boundary conditions. Consequently, the resulting stress-energy tensor captures precisely how the detector's presence affects the spacetime geometry through the semiclassical backreaction, while the bound state associated with the detector itself remains dynamically decoupled from the \victorr{backreaction effects}.

The evaluation of the renormalized stress-energy tensor proceeds via the point-splitting regularization procedure \cite{decanini2008hadamard}. In this approach, the expectation value of the stress-energy tensor operator for a massive and minimally couple scalar field is defined through the regularized expression \begin{multline}
    \left\langle T_{\mu\nu}\right\rangle_{\textrm{ren}} =\lim_{x\to x'} \left[ \left(g_{\nu}^{\ \nu'}\nabla_{\mu}\nabla_{\nu'}-\frac{1}{2}g_{\mu\nu}g^{\rho\sigma'}\nabla_{\rho}\nabla_{\sigma'}\right.  +\right. \\ \left.
    \left.-\frac{1}{2}g_{\mu\nu}m^{2}  \right)G_{\textrm{ren}}\left(x,x'\right)\right],
\end{multline} where $G_{\textrm{ren}}(x,x')$ corresponds to the complete Green's function after the Hadamard subtraction. The bivector of parallel transport from $x$ to $x'$ is denoted by $g_{\mu \nu'}$ and it is defined by the differential equation $\nabla_\rho g_{\mu \nu '}\nabla^\rho \sigma =0$ and the condition $g_{\mu \nu'} (x,x')=g_{\mu \nu }(x)$ when $x'\to x$. Hence, by direct computation, the resulting nonvanishing components are \begin{multline}
   \left\langle T_{tt}\right\rangle_{\textrm{ren}} =\frac{m_{0}}{4\pi^{2}r^{3}}\int_{0}^{\infty}\frac{ds}{\beta}\cosh\left(\frac{s}{\beta}\right)K_{1}\left(m_{0}\left(2r+s\right)\right)+\\+\left(\frac{1}{8\pi^{2}r^{4}}+\frac{m_{0}^{2}}{4\pi^{2}r^{2}}\right)\int_{0}^{\infty}\frac{ds}{\beta}\cosh\left(\frac{s}{\beta}\right)K_{0}\left(m_{0}\left(2r+s\right)\right),
\end{multline} 

\begin{multline}
   \left\langle T_{rr}\right\rangle_{\textrm{ren}} = \frac{1}{8\pi^{2}r^{4}}\int_{0}^{\infty}\frac{ds}{\beta}\cosh\left(\frac{s}{\beta}\right)K_{0}\left(m_{0}\left(2r+s\right)\right)+ \\+\frac{m_{0}}{4\pi^{2}r^{3}}\int_{0}^{\infty}\frac{ds}{\beta}\cosh\left(\frac{s}{\beta}\right)K_{1}\left(m_{0}\left(2r+s\right)\right),
\end{multline} 

\begin{multline}
    \left\langle T_{\theta\theta}\right\rangle_\textrm{ren} =-\frac{m_{0}}{4\pi^{2}r}\int_{0}^{\infty}\frac{ds}{\beta}\cosh\left(\frac{s}{\beta}\right)K_{1}\left(m_{0}\left(2r+s\right)\right)\\-\frac{m_{0}^{2}}{8\pi^{2}}\int_{0}^{\infty}\frac{ds}{\beta}\cosh\left(\frac{s}{\beta}\right)K_{2}\left(m_{0}\left(2r+s\right)\right)\\-\left(\frac{1}{8\pi^{2}r^{2}}+\frac{m_{0}^{2}}{8\pi^{2}}\right)\int_{0}^{\infty}\frac{ds}{\beta}\cosh\left(\frac{s}{\beta}\right)K_{0}\left(m_{0}\left(2r+s\right)\right),
\end{multline} and $\langle T_{\phi \phi }\rangle_\textrm{ren}=\sin^2(\theta) \left\langle T_{\theta\theta}\right\rangle_\textrm{ren}$. It is important to note that the contributions to the stress-energy tensor arise exclusively from $G_\textrm{bound}(x,x')+G_\beta (x,x')$, as the Dirichlet sector leads to a vanishing contribution. Furthermore, since the bound-state mode is exactly canceled in the total Green's function, the detector itself does not contribute to the energy-density content of the quantum field \victorr{for the vacuum state}~\footnote{\victorr{We stress that we have verified this cancellation of the bound state contribution only in the vacuum state. One could, in principle, consider the contributions of the detector exited states to $\langle T_{\mu \nu} \rangle$ and its correspondent backreaction effects on spacetime, but these investigations are beyond the scope of this manuscript.}}. A crucial property of the obtained tensor is that it satisfies the conservation law $\nabla_{\mu} \langle T^{\mu\nu} \rangle_{\textrm{ren}} = 0$, as guaranteed by the point-splitting regularization procedure when properly implemented.

\begin{figure}[h!]
    \centering
    \includegraphics[width=0.47\textwidth]{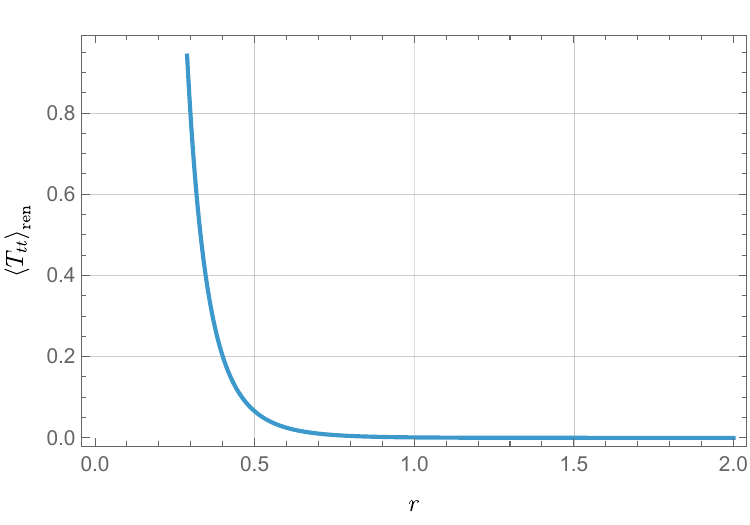}
    \caption{Renormalized energy density $\langle T_{tt}\rangle_{\textrm{ren}}$ for $m_{0}=2$ and $\beta=1$. 
    The boundary contribution dominates near $r=0$ and rapidly decays as $r$ increases.}
    \label{fig:Ttt}
\end{figure}

\begin{figure}[h!]
    \centering
    \includegraphics[width=0.47\textwidth]{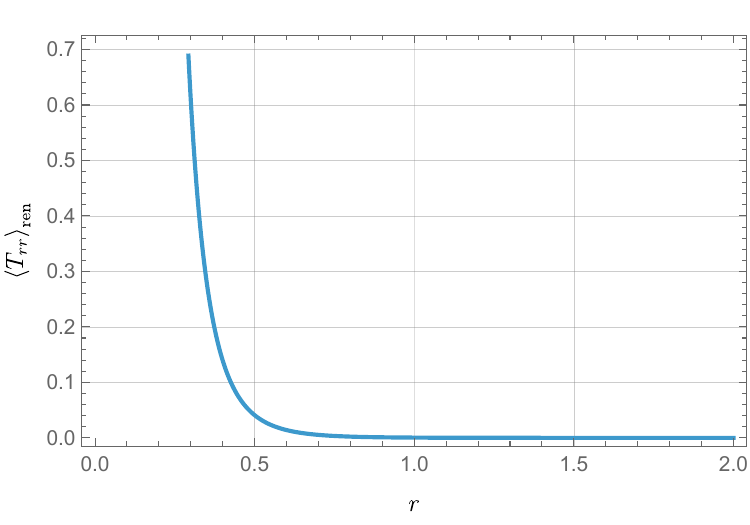}
    \caption{Radial component $\langle T_{rr}\rangle_{\textrm{ren}}$ for $m_{0}=2$ and $\beta=1$. 
    The same localized boundary effect near $r=0$ is observed, with fast decay at larger distances.}
    \label{fig:Trr}
\end{figure}

\begin{figure}[h!]
    \centering
    \includegraphics[width=0.47\textwidth]{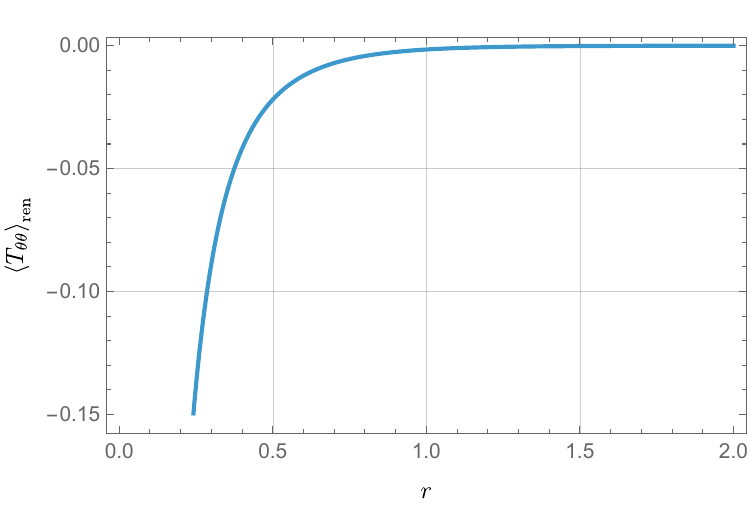}
    \caption{Angular component $\langle T_{\theta\theta}\rangle_{\textrm{ren}}$ for $m_{0}=2$ and $\beta=1$. 
    It also exhibits the localized enhancement close to $r=0$ and tends to vanish quickly away from the excised point. 
    The relation $\langle T_{\phi\phi}\rangle_{\textrm{ren}}=\sin^2(\theta)\,\langle T_{\theta\theta}\rangle_{\textrm{ren}}$ follows straightforwardly.}
    \label{fig:Tthetatheta}
\end{figure}

To complement this analysis, in Figs.~\ref{fig:Ttt}--\ref{fig:Tthetatheta} we show the behavior of the non-vanishing components of the renormalized stress-energy tensor for the illustrative choice $m_{0}=2$ and $\beta=1$. As discussed above, the Dirichlet sector does not contribute, and the entire result comes from the boundary-dependent part. The plots reveal that all components are strongly influenced by the presence of the excised point $r=0$: near this region, the boundary-induced contributions dominate, while for increasing $r$ they decay rapidly, leaving no residual contribution at large distances. This behavior highlights the fact that the effect of the boundary conditions is highly localized around $r=0$, in contrast with the constant role played by the Dirichlet term in $\langle \Psi^{2}\rangle$.

\section{Concluding Remarks}
\label{concluding remarks}

In this work we have introduced a fully relativistic detector model formulated as a quantum field in Minkowski spacetime with the spatial origin excised. By imposing Robin-type boundary conditions at the puncture, we showed that the resulting spectral structure naturally contains discrete bound modes, which play the role of localized detector degrees of freedom. Unlike models that require ad hoc confining potentials, here the detector’s internal structure emerges intrinsically from the self-adjoint extensions of the radial operator.

We constructed the two-point function of the detector field and demonstrated its decomposition into three distinct contributions: the discrete bound-state sector, the modified continuum, and the Dirichlet sector. A remarkable feature of this decomposition is the exact cancellation of the discrete contribution in the renormalized observables, leaving only boundary-induced effects. The evaluation of the stress–energy tensor confirmed this result, showing that the detector itself carries no net stress-energy, while the boundary conditions are solely responsible for the localized modifications of the vacuum.

Our formulation provides a covariant, field-theoretic framework that reproduces the Unruh-DeWitt detector response at leading order, but without the conceptual limitations of introducing nonrelativistic degrees of freedom. This establishes a direct bridge between traditional detector models and localized field-based detectors, clarifying the physical origin of discrete localized modes. 

We stress that the artificial singularity at $r=0$ in our model acquires a genuine physical interpretation when a quantum field is considered in the background of a global monopole~\cite{pitelliglobalmonpole}. In that setting, $r=0$ corresponds to a true curvature singularity, and the self-adjoint extension framework developed here applies in close analogy except that, in contrast to the present case, the Dirichlet sector yields a nonvanishing contribution.

Natural extensions of this work include the analysis of detectors coupled through compactly supported interactions, the study of responses in nonvacuum states, and generalizations to curved backgrounds. In particular, applying this framework to spacetimes with horizons may provide new insights into particle detection in gravitational settings, while maintaining a \victor{covariant} and self-consistent description.

\section*{Acknowledgments}

V.H.M.R. acknowledges the financial support of Coordenação de Aperfeiçoamento de Pessoal de Nível Superior (CAPES) – Brazil, Finance Code 001, and gratefully acknowledges the kind hospitality of the Department of Mathematics of the University of Genova. J. P. M. P. thanks the support provided in part by Conselho Nacional de Desenvolvimento Científico e Tecnológico (CNPq, Brazil), Grant No. 305194/2025-9.


\begin{thebibliography}{0}%
\makeatletter
\providecommand \@ifxundefined [1]{%
 \@ifx{#1\undefined}
}%
\providecommand \@ifnum [1]{%
 \ifnum #1\expandafter \@firstoftwo
 \else \expandafter \@secondoftwo
 \fi
}%
\providecommand \@ifx [1]{%
 \ifx #1\expandafter \@firstoftwo
 \else \expandafter \@secondoftwo
 \fi
}%
\providecommand \natexlab [1]{#1}%
\providecommand \enquote  [1]{``#1''}%
\providecommand \bibnamefont  [1]{#1}%
\providecommand \bibfnamefont [1]{#1}%
\providecommand \citenamefont [1]{#1}%
\providecommand \href@noop [0]{\@secondoftwo}%
\providecommand \href [0]{\begingroup \@sanitize@url \@href}%
\providecommand \@href[1]{\@@startlink{#1}\@@href}%
\providecommand \@@href[1]{\endgroup#1\@@endlink}%
\providecommand \@sanitize@url [0]{\catcode `\\12\catcode `\$12\catcode `\&12\catcode `\#12\catcode `\^12\catcode `\_12\catcode `\%12\relax}%
\providecommand \@@startlink[1]{}%
\providecommand \@@endlink[0]{}%
\providecommand \url  [0]{\begingroup\@sanitize@url \@url }%
\providecommand \@url [1]{\endgroup\@href {#1}{\urlprefix }}%
\providecommand \urlprefix  [0]{URL }%
\providecommand \Eprint [0]{\href }%
\providecommand \doibase [0]{http://dx.doi.org/}%
\providecommand \selectlanguage [0]{\@gobble}%
\providecommand \bibinfo  [0]{\@secondoftwo}%
\providecommand \bibfield  [0]{\@secondoftwo}%
\providecommand \translation [1]{[#1]}%
\providecommand \BibitemOpen [0]{}%
\providecommand \bibitemStop [0]{}%
\providecommand \bibitemNoStop [0]{.\EOS\space}%
\providecommand \EOS [0]{\spacefactor3000\relax}%
\providecommand \BibitemShut  [1]{\csname bibitem#1\endcsname}%
\let\auto@bib@innerbib\@empty
\end{thebibliography}%


\begin{thebibliography}{99}
\bibitem{unruh1976notes}
 W. G. Unruh, {\it Notes on black-hole evaporation}, Phys. Rev. D {\bf 14}, 870 (1976).

\bibitem{dewitt1979general}
B. DeWitt, {\it General Relativity: An Einstein Centenary Survey} (Cambridge University Press, Cambridge, England, 1979).

\bibitem{crispino2008unruh}
L. C. B. Crispino, A. Higuchi, and G. E. A. Matsas, {\it The Unruh effect and its applications}, Rev. Mod.  Phys. {\bf 80}, 787 (2008).

 \bibitem{liu2022does} 
 Z. Liu, J. Zhang,  R. B. Mann, and H. Yu, {\it Does acceleration assist entanglement harvesting?}, Phys. Rev. D {\bf 105}, 085012 (2022).

\bibitem{APozas-Kerstjens}
A. Pozas-Kerstjens and E. Mart\'in-Mart\'inez, {\it Harvesting
correlations from the quantum vacuum}, Phys. Rev. D {\bf 92},
064042 (2015).

\bibitem{Foo}
J. Foo, R. B. Mann, and M. Zych, {\it Entanglement amplification between superposed detectors in flat and curved
spacetimes}, Phys. Rev. D {\bf 103}, 065013 (2021).

\bibitem{de2021relativistic}
J. de Ram{\'o}n, M. Papageorgiou, and E. Mart{\'\i}n-Mart{\'\i}nez, {\it Relativistic causality in particle detector models: Faster-than-light signaling and impossible measurements}, Phys. Rev. D {\bf 103}, 085002 (2021).

\bibitem{martin2015causality}
E. Mart{\'\i}n-Mart{\'\i}nez, {\it Causality issues of particle detector models in QFT and quantum optics}, Phys. Rev. D {\bf 92}, 104019 (2015).

\bibitem{martin2021broken}
E. Mart{\'\i}n-Mart{\'\i}nez, T. R. Perche, and B. S. L. Torres, {\it Broken covariance of particle detector models in relativistic quantum information}, Phys. Rev. D {\bf 103}, 025007 (2017).

\bibitem{de2023causality}
J. de Ram{\'o}n, M.  Papageorgiou, and E. Mart{\'\i}n-Mart{\'\i}nez, {\it Causality and signalling in noncompact detector-field interactions}, Phys. Rev. D {\bf 108}, 045015 (2023).

\bibitem{fewster2020quantum}
C. J. Fewster and R. Verch, {\it Quantum fields and local measurements}, Commun. Math. Phys.  {\bf 378}, 851 (2020).

\bibitem{fewster2020generally}
C. J. Fewster, {\it  Generally covariant measurement scheme for quantum field theory in curved spacetimes}, Progress and Visions in Quantum Theory in View of Gravity: Bridging foundations of physics and mathematics, 253 (2020).

\bibitem{perche2024particle}
T. R. Perche, J. Polo-G\'omez, B. S. L. Torres and E. Mart{\'\i}n-Mart{\'\i}nez, {\it Particle detectors from localized quantum field theories}, Phys. Rev. D {\bf 109}, 045013 (2024).

\bibitem{bruno1}
B. de S. L. Torres, {\it Particle detector models from path integrals of localized quantum fields}, Phys. Rev. D {\bf 109}, 065004 (2024).

\bibitem{bruno2}
B. Ragula, B. de S. L. Torres, E. Schnetter, and E. Mart{\'\i}n-Mart{\'\i}nez, {\it Localizing quantum fields with time-dependent potentials}, Phys. Rev. D {\bf 111}, 105029 (2025).

\bibitem{kay}
B.~S.~Kay and U.~M.~Studer,
{\it Boundary conditions for quantum mechanics on cones and fields around cosmic strings},
Commun. Math. Phys. \textbf{139}, 103 (1991).

\bibitem{pitelliglobalmonpole}
J. P. M. Pitelli and P. S. Letelier, {\it Quantum singularities around a global monopole}, Phys. Rev. D {\bf 80}, 104035 (2009).

\bibitem{gitman2010self}
D. M. Gitman, I. V. Tyutin and B. L. Voronov, {\it Self-adjoint extensions and spectral analysis in the Calogero problem},  J. Phys. A {\bf 43}, 145205 (2010).

\bibitem{WaldII}
A.~Ishibashi and R.~M.~Wald,
{\it Dynamics in non-globally-hyperbolic static spacetimes II: General analysis of prescriptions for dynamics},
Class.\ Quantum Grav. \textbf{20}, 3815 (2003). [arXiv:gr-qc/0305012].

\bibitem{perche2025stress}
T. R. Perche, J. P. M. Pitelli and D. A. T. Vanzella, {\it Stress-energy tensor of an Unruh-DeWitt detector}, Phys. Rev. D {\bf 111}, 045004 (2025).

\bibitem{blinder1988green}
S. M. Blinder, {\it Green’s function and propagator for the one-dimensional $\delta$-function potential}, Phys. Rev. A {\bf 37}, 973 (1988).

\bibitem{blinder1981nonrelativistic}
S. M. Blinder, {\it Nonrelativistic Coulomb Green’s function in parabolic coordinates}, J. Math. Phys. (N.Y) {\bf 22}, 306 (1981).

\bibitem{hostler1964coulomb}
L. Hostler, {\it Coulomb Green's functions and the Furry approximation} J. Math. Phys. (N.Y.) {\bf 5}, 591 (1964).

 \bibitem{decanini2008hadamard}
 Y. Décanini and A. Folacci, {\it Hadamard renormalization of the stress-energy tensor for a quantized scalar field in a general spacetime of arbitrary dimension}, Phys. Rev. D {\bf 78}, 044025 (2008).

\bibitem{Gradshteyn}
I. Gradshteyn and I. Ryzhik, {\it Table of Integrals, Series and Products (Corrected and Enlarged Edition Prepared by D. Zwillinger)} (Academic Press, New York, 2014).

\end{thebibliography}
\end{document}